\documentclass[authoryear]{elsarticle}
\usepackage[T1]{fontenc}
\usepackage[margin=1.5in]{geometry}
\usepackage{graphicx} 
\usepackage{subcaption}
\usepackage{amssymb}
\usepackage{amsmath}
\usepackage{amsfonts}
\usepackage{multirow}
\usepackage{tabularray}
\usepackage{natbib}
\setcitestyle{authoryear,round}
\usepackage{url}

\usepackage{tabularray}

\makeatletter
\def\ps@pprintTitle{%
    \let\@oddhead\@empty
    \let\@evenhead\@empty
    \def\@oddfoot{\footnotesize\itshape
         {} \hfill\today}%
    \let\@evenfoot\@oddfoot
    }
\makeatother

\begin{document}

\title{Impact of distribution fees on BESS scheduling and profitability}

\begin{abstract}

Battery energy storage systems (BESS) are expected to play an important role in electricity markets with increasing shares of renewable generation. While existing research has primarily focused on price arbitrage and ancillary services, the role of grid fees in shaping BESS operation and profitability remains insufficiently understood. This article investigates how different levels of distribution fees affect the scheduling and economic viability of BESS in the day-ahead electricity market.

The analysis employs a mixed-integer linear programming model of BESS operation combined with electricity price data from the German market. Four system configurations are considered: stand-alone storage and BESS combined with consumption, generation, or both. The value of storage is measured as the difference between system profits with and without BESS. In addition, a rolling-horizon optimization framework is used to evaluate the impact of forecast uncertainty and decision horizon length on operational outcomes.

The results show that grid fees significantly influence both BESS profitability and operational strategies. For stand-alone storage, higher transmission charges reduce arbitrage revenues and battery utilization. When BESS is integrated with consumption and generation units, load shifting and self-consumption become the dominant sources of value, leading to a non-monotonic relationship between grid fees and storage profitability. 
These findings highlight the importance of considering tariff structures when evaluating storage investments and designing regulatory frameworks for electricity markets with increasing flexibility needs.




\end{abstract}

\author[1]{Katarzyna Maciejowska\corref{cor1}}
\ead{katarzyna.maciejowska@pwr.edu.pl}

\affiliation[1]{organization={Department of Operations Research and Business Intelligence, Wrocław University of Science and Technology}, 
addressline={Wybrzeże Wyspiańskiego 27}, 
postcode={50-370},
city={Wrocław}, 
country={Poland}}

\cortext[cor1]{Corresponding author}

\begin{keyword}
  electricity market \sep battery energy storage system
\end{keyword}

\maketitle

\section{Introduction}

The Net Zero Emissions by 2050 Scenario foresees a substantial expansion of renewable energy sources, such as solar PV and wind. This expansion is accompanied by a significant rise in electricity demand, driven by the electrification of end-use sectors. In this context, grid-scale energy storage -- particularly battery systems -- will play a critical role in mitigating grid impacts. It will also support the management of both short-term and seasonal variability in renewable generation \citep{chat:etal:24}. This is essential to ensure system stability and reliability under increasing demand.

Energy storage systems (ESS) enable the temporal shifting of electricity by storing surplus energy and dispatching it during periods of higher system need, such as low solar availability or grid disturbances caused by extreme weather events. Pumped-storage hydropower (PSH) is currently the most mature and widely deployed storage technology, with a worldwide installed capacity of around 169~GW in 2021 \citep{eia:23}. However, its further expansion is constrained by site-specific geographical requirements.At the same time, battery energy storage systems (BESS) are emerging as a key alternative due to their modularity, flexibility, and wide range of deployable capacities. Although their installed capacity remains smaller than that of PSH, grid-scale batteries are projected to account for the majority of future storage growth worldwide. They are typically used for sub-hourly, hourly, and daily balancing applications. Total installed grid-scale battery capacity increased from nearly 52~GW in 2022 to 80~GW by the end of 2024 \citep{eia:23, iea:25}, reflecting rapid growth in recent years.

The development of battery energy storage systems (BESS) depends on market regulations and economic factors that directly affect investment profitability. The economic viability of BESS is primarily driven by four revenue streams: price arbitrage, load shifting, frequency regulation, and voltage support \citep{yam:etal:22, sch:sta:23, sha:26}. The literature has focused predominantly on price arbitrage \citep{mer:etal:23, degu:etal:25}, as it represents a natural application of grid-scale BESS. 
\cite{yam:etal:22, louk:etal:21} also discuss revenues from providing additional services, such as regulation services, peak shaving, and load shifting. Their findings indicate that combining multiple revenue streams can substantially increase the net present value of BESS. However, participation in certain ancillary services may require multiple licenses, which increases costs and can hinder deployment.

The profitability of BESS also depends on the costs of construction and operation. This study focuses on short-term operational decisions of BESS, which are primarily influenced by variable costs, among which energy costs play a crucial role. In the literature, these costs are typically approximated by day-ahead (DA) electricity prices, which serve as reference prices for many power purchase agreements. Although widely used, DA prices underestimate actual energy costs because they do not include distribution fees charged for using the grid to transport electricity. Moreover, BESS operate as buyers and sellers and therefore are subject to grid fees applied to both generators and consumers.  According to the report by \cite{Bundes:24}, distribution fees in Germany in 2024 amounted to 11.62~ct/kWh for households, 9.42~ct/kWh for commercial consumers, and 4.12~ct/kWh for energy-intensive industries. Over the same period, the average day-ahead electricity market price was approximately 92.88~EUR/MWh. Hence, grid fees constitute a significant component of the total cost of electricity purchases and may represent a barrier to BESS deployment \citep{yan:etal:18}.


The literature examines the relationship between BESS and grid fees from two main perspectives. First, the design of grid fee structures is analyzed, as in \cite{degu:etal:25}. That study compares flat and dynamic tariffs and evaluates their impact on BESS performance. The analysis assumes symmetric fees applied to both electricity purchases and sales. However, in most countries, grid fees are imposed only on electricity consumption \citep{ENTSO:23}. Moreover, the proposed dynamic tariffs do not fully reflect actual market conditions, as they assume that fees increase with electricity prices. In practice, the opposite structure often applies: low electricity prices incentivize higher consumption, which in turn leads to higher transmission costs. 

\cite{ber:etal:26} consider more realistic scenarios, in which grid fees -- both energy- and power-based -- are either fixed, as in current regulations, or the energy component depends on electricity prices. To evaluate the impact of dynamic tariffs, they use a dataset describing actual load profiles of 50 representative companies in Germany. The study employs mixed-integer linear programming (MILP)  to optimize battery operation and sizing using electricity price data from 2024. 
The analysis focuses on peak shaving and the relationship between reductions in maximum power consumption and decreases in energy procurement and grid usage costs. The results confirm that the primary driver of BESS value is the use of storage for load management rather than price arbitrage in the electricity market. 
Although the article proposes and evaluates new types of dynamic tariffs, it does not analyze how adjustments to existing fixed tariffs affect BESS operation and economic viability. Moreover, the study considers BESS only in combination with load and does not account for the presence of generation units, which may significantly influence the profitability of battery installations.


The second stream of the literature examines how existing transmission tariffs affect the profitability of BESS. \cite{ade:etal:2018} studies battery systems engaged in price arbitrage and shows that network charges can alter battery dispatch decisions and render many marginal arbitrage cycles unprofitable. This, in turn, reduces battery utilization and decreases overall BESS revenues. \cite{mer:etal:23} evaluates arbitrage opportunities across European countries. A case study of Belgium incorporates grid fees that differ between charging and discharging. However, the paper considers a single set of transmission fees and therefore does not allow for an assessment of how fee levels affect the profitability of BESS. 
Finally, \cite{tiem:etal:20} considers a comprehensive dataset that includes information on more than 5{,}000 companies and the grid fees they face. The article focuses on power-based charges and analyzes the value of BESS from the perspective of peak shaving.

Despite the importance of grid fees for the operational economics of energy storage, their impact on BESS scheduling and profitability remains insufficiently explored. In particular, the literature lacks analyses examining how changes in grid fee levels affect the viability and operation of storage systems. This issue is becoming increasingly important in light of recent regulatory changes observed in Germany. Moreover, the problem becomes complex when BESS is combined with either consumption or generation units \citep{louk:etal:21, yam:etal:22, jiv:etal:25, chat:etal:24}. In such cases, the value of the battery may arise primarily from load management, as shown in \cite{ber:etal:26}, potentially resulting in a non-monotonic relationship between grid fees and BESS gains.

This article addresses this gap by investigating how different levels of the energy-based component of grid fees affect the operation and value of BESS. In this study, the size of BESS is assumed to be predefined and fixed; therefore, gains from reductions in maximum load are not included in the analysis. Instead, four scenarios are considered: a stand-alone battery and a BESS combined with a generation unit, a consumption unit, or both.

The value of the battery is measured as the difference between system profits with and without BESS. For a stand-alone BESS, its value corresponds to the profit obtained from price arbitrage. When the battery is combined with a consumption unit, the value added by BESS consists of reduced electricity purchase costs due to load shifting, in addition to arbitrage profits. Finally, when BESS operates alongside both generation and consumption units, three sources of additional value arise: price arbitrage, shifting of electricity sales and purchases over time, and reduced grid fee payments through increased self-consumption.


The operation of BESS is described using a power--energy model \citep{hut:etal:25}, optimized with a mixed-integer linear programming approach as in \cite{mor:etal:15}, \cite{ade:etal:2018} and \cite{mer:etal:23}. The algorithm determines the BESS schedule in order to maximize its value while accounting for the physical limitations of battery operation. In the model, the value of the battery is measured as the difference between system profits or energy procurement costs with and without BESS. 
The electricity cost is based on the market data from the German day-ahead market over the period 01.01.2022--31.12.2024.

This article contributes to the literature in the following areas:
\begin{itemize}
\itemsep = 0pt
    \item We analyze how grid fees affect the operational viability of BESS engaged in price arbitrage and load management. A range of transmission fee levels is considered, corresponding to current market conditions and reflecting ongoing regulatory changes.  
    
    \item Four scenarios are examined in which BESS either operates as a stand-alone unit or is combined with consumption or generation units. As a result, three sources of battery value arise: price arbitrage, intertemporal shifting of electricity trade, and a reduction in grid fee payments through increased self-consumption. The results show that load management and the reduction of transmission costs constitute the primary sources of BESS value.
    
    \item The results confirm that higher transmission fees reduce income from price arbitrage. However, when BESS is combined with both generation and consumption units, the relationship between grid fees and the value added by BESS becomes non-monotonic. As a result, BESS value may increase with higher transmission charges due to a greater potential for reducing transmission payments through load shifting and self-consumption.
    
    \item The study employs a rolling window framework. The findings indicate that extending the decision horizon from one day to two days improves BESS scheduling outcomes. However, further extending the horizon to three days has only a minor impact on financial performance while increasing computational complexity. This result supports the use of a relatively short, two-day decision horizon in the optimization algorithm.
\end{itemize}

The paper is structured as follows. Section \ref{sec:2} introduces the modeling framework, Section \ref{sec:results} presents and discuss the results of the analysis. Finally, Section \ref{sec:conclusions} concludes.

\section{Modeling framework}\label{sec:2}

\begin{table}
\centering
\caption{Notation}
\label{tab:notation}
 \begin{tblr}{
  colspec={c|l|c},
  cell{2,17}{1} = {c=3}{c}
}
\hline
Notation & Description & Units \\
\hline
BESS & &\\
\hline
$\eta$ & charging/discharging efficiency & \%\\
$H$ & time needed to fully charge/discharge the battery & h\\
\hline
$E^{max}$ & the maximum energy charged/discharged within an hour & MWh \\
$E_{t,h}^{ch}$ & energy charged within an hour & MWh\\
$E_{t,h}^{dis}$ & energy discharged within an hour & MWh\\
\hline
$S^{max}$ & net energy capacity rating & MWh\\
$SOC_0$ & initial state of charge of the battery &MWh\\
$SOC_{t,h} $ & state of charge of the battery: $SOC_0 + \sum_{k=1}^h (E_{t,k}^{ch}+E_{t,k}^{dis})$ & MWh\\
\hline
$G_{t,h}$ & generation &MWh\\
$C_{t,h}$ & self-consumption &MWh\\
$NG_{t,h}$ & net generation: $G_{t,h}-C_{t,h}$ &MWh\\
$X_{t,h}$ & exchange with the grid:  $NG_{t,h} +E_{t,h}^{dis}\eta +  E_{t,h}^{ch}/\eta$ &MWh\\
$X_{t,h}^{(+)}$ & sale to the grid:  $X_{t,h}1_{X_{t,h}\geq 0}$ &MWh\\
$X_{t,h}^{(-)}$ & purchase from the grid:  $X_{t,h}1_{X_{t,h}< 0}$ &MWh\\
\hline
Market & &\\
\hline
$P_{t,h}$ & electricity price & EUR/MWh\\
$\hat{P}_{t,h}$ & forecasted electricity price &EUR/MWh \\
$F$ & distribution fee &ct/kWh\\
\hline
\end{tblr}

\vspace{0.3cm}
\small{Notice: Indices $t$ and $h$ denote a day and an hour, respectively. The maximum energy charged/discharged within an hour can be computed as $E^{max}=S^{max}/H$}
\end{table}

In this study, a battery energy storage system  that participates in the day-ahead (DA) electricity market is considered. Unlike many existing analyses, we model BESS as a part of a broader system that may include both generation and consumption units. In the remainder of this paper, it is assumed that these units are co-located and the proximity is sufficient to avoid the use of the grid. Consequently, energy transfers between the BESS and the associated generation or consumption units are not subject to distribution fees.

The performance of a BESS depends strongly on its technical characteristics. Here\, these are summarized by the net energy capacity  ($S^{max}$), the time required to fully charge or discharge ($H$), and the charging and discharging efficiency ($\eta$). The energy capacity represents the maximum amount of energy that the BESS can store. In this analysis, it is expressed as the net capacity, defined as the difference between the maximum allowable state of charge and the minimum energy level required to prolong battery lifetime and mitigate degradation. Batteries also differ in the time required to reach full charge or discharge, $H$.
In the literature, this property is often expressed as the C-rating, which is the inverse of $H$ and indicates the fraction of total capacity that can be charged or discharged in one hour. For example, a battery that can reach full capacity in one hour has $H=1h$ and a C-rating of 1. If two hours are required ($H=2h$), the corresponding C-rating is 0.5. Thus, the longer the charging time, the lower the C-rating. This characteristic significantly affects operational performance because it reflects the maximum amount of energy that can be charged or discharged within an hour ($E^{max}$). Finally, BESS operation involves energy losses during charging and discharging. New systems may exhibit efficiencies close to unity, but these typically decline by 1–2\% per year. After five years of operation, this corresponds to a 5–10\% reduction in usable capacity compared to initial performance. This gradual degradation is expected; however, its rate depends on several factors, including the frequency of charge–discharge cycles, the depth of discharge, and environmental conditions.

The system analyzed in this work is additionally characterized by generation ($G_{t,h}$) and consumption ($C_{t,h}$) profiles that may vary across days ($t$) and hours ($h$). Here, in order to simplify the analysis, stylized generation and consumption patterns are applied; however, the proposed procedure can be directly applied to more realistic schedules. Finally, it is assumed that the entire  system participates in the DA market, submitting offers before noon on the day preceding delivery. As a result, all BESS scheduling decisions -- specifically, when and how much energy to charge or discharge -- are made before actual prices ($P_{t,h}$) are known and therefore rely on price forecasts.

When the system exchanges energy with the grid, it is subject to an additional distribution fee~($F$), charged per MWh of electricity purchased. As a result, the buying and selling prices of electricity differ. Specifically, when electricity is injected into the grid, the system receives the market price $P_{t,h}$, whereas when electricity is withdrawn from the grid, it pays $P_{t,h} + F$. Because the BESS may act both as a generator when discharging (selling electricity) and as a consumer when charging (purchasing electricity), the grid fee has a substantial impact on its overall performance and operational profitability, as we will show in the analysis.

\subsection{BESS scheduling}

The objective function in MILP is the profit from electricity generation and price arbitrage, or a costs of electricity purchases needed to cover the demand. The income (or cost) of operating the system on day~$t$ can be expressed as
\begin{equation}\label{eq:profit}
    \pi_{t}^{BESS} = \sum_{h=1}^{24} P_{t,h} X_{t,h}^{(+)} 
    + \sum_{h=1}^{24} \left(P_{t,h} + F\right) X_{t,h}^{(-)},
\end{equation}
where $P_{t,h}$ represents the market price on day $t$ and hour $h$ and $F$ is the grid fee. 
The variable $X_{t,h}$ denotes the net exchange with the grid and is defined as
\begin{equation}
    X_{t,h} = NG_{t,h} + E_{t,h}^{\mathrm{dis}} \eta 
    + \frac{E_{t,h}^{\mathrm{ch}}}{\eta},
\end{equation}
where $NG_{t,h} = G_{t,h} - C_{t,h}$ is the net generation at hour~$h$, and 
$E_{t,h}^{\mathrm{ch}}$ and $E_{t,h}^{\mathrm{dis}}$ denote the amounts of energy charged into or discharged from the BESS, respectively. Notation and description of variables used in the decision procedure are summarized by Table \ref{tab:notation}.
The terms $X_{t,h}^{(+)}$ and $X_{t,h}^{(-)}$ in (\ref{eq:profit}) represent, respectively, the amount of energy sold to the grid at hour~$h$ and the amount purchased from the grid. The system is not allowed to buy and sell electricity simultaneously within the same hour; therefore, one of the two values must equal zero for any given~$h$. Consistent with their interpretation as outgoing and incoming energy flows, the variables satisfy
\[
    X_{t,h}^{(+)} \ge 0, \qquad X_{t,h}^{(-)} < 0.
\]

The value of $\pi_{t}$ depends on the traded quantities, the market prices, and the grid fee $F$. However, at the time of scheduling, the BESS does not yet know the following day's prices. Instead, decisions rely on forecasted revenue, given by
\begin{equation}\label{eq:avg_profit_forecast}
    \hat{\pi}_{t}^{BESS} = \sum_{h=1}^{24} \hat{P}_{t,h} X_{t,h}^{(+)} 
    + \sum_{h=1}^{24} \left(\hat{P}_{t,h} + F\right) X_{t,h}^{(-)},
\end{equation}
where the market prices $P_{t,h}$ are replaced by their forecasts $\hat{P}_{t,h}$.

Due to technical constraints, the operation of the battery is subject to the following conditions:
\begin{equation}\label{eq:4}
  0 \leq E_{t,h}^{\mathrm{dis}} \leq b_{t,h}E^{\max},
\end{equation}
\begin{equation}\label{eq:5}
  -(1-b_{t,h})E^{\max} \leq E_{t,h}^{\mathrm{ch}} \leq 0.
\end{equation}
The inequalities, (\ref{eq:4})--(\ref{eq:5}), state that the energy discharged from or charged into the battery is bounded by the maximum hourly transfer capability $E^{\max}$. The binary  variable $b_{t,h}\in\{0,1\}$ ensures that it is
impossible to charge and discharge the battery at the same time. 

Additionally, the minimum and maximum state of charge, $SOC_{t,h}$, are bounded:
\begin{equation}\label{eq:6}
  0 \leq SOC_{t,h} \leq S^{\max}.
\end{equation}
The state of charge is determined by the initial battery level and the energy inflows and outflows during the considered periods. Suppose that BESS makes decisions on day $d-1$ regarding charging and discharging hours on day $d$. In this case, the initial state of charge is given by $SOC_0 = SOC_{d-1,24}$ and varies from day to day.

Although the battery schedules its operation on a daily basis, it may consider a longer time horizon when optimizing its decisions (see \cite{mer:etal:23} for a discussion). Let $D$ denote the number of consecutive days considered by BESS when determining charging and discharging schedules. We assume that, at the end of the decision horizon, BESS remains half charged:
\begin{equation}\label{eq:7}
  SOC_{d+D,24} = 0.5 S^{\max}.
\end{equation}
It should be emphasized that even when the optimization horizon spans multiple days, BESS uses charging and discharging decisions only for day $d$. After each day, the battery updates price information and re-optimizes its operation using newly available forecasts.

Finally, as the time horizon $D$ increases, the impact of constraint (\ref{eq:7}) on the schedule for day $d$ becomes weaker. The choice of $D$ is, however, limited by the availability of price forecasts used in (\ref{eq:avg_profit_forecast}). \cite{mer:etal:23} uses 7 days horizons.  In this research, we set $D = 2$ as the baseline specification and examine how scheduling decisions depend on the choice of the decision horizon.


\subsection{Evaluation of BESS performance}

The performance of BESS can be assessed from several perspectives. In this work, we consider three key measures. The first is the average daily profit resulting from the integration of the BESS into the system. It is computed as the difference between the daily profit of the system with BESS (defined in~\eqref{eq:profit}) and the corresponding money flow of the same system -- with identical generation and consumption profiles -- but operating without a BESS. In the absence of storage, the daily profit on day~$t$ is given by
\begin{equation}\label{eq:profit_no_BESS}
    \pi_{t}^{\mathrm{no\,BESS}} 
    = \sum_{h=1}^{24} P_{t,h}\, NG_{t,h} \mathbf{1}_{\{NG_{t,h}>0\}}
      + \sum_{h=1}^{24} \left(P_{t,h} + F\right) NG_{t,h}\mathbf{1}_{\{NG_{t,h}<0\}},
\end{equation}
where $\mathbf{1}_{\{NG_{t,h}>0\}}$ and $\mathbf{1}_{\{NG_{t,h}<0\}}$ are indicator functions equal to~1 when the respective conditions are satisfied, and~0 otherwise.

The average additional added value from deploying the BESS, normalized per 1~MWh of net capacity, is defined as
\begin{equation}\label{eq:avg_profit_per_MWh}
    V = \frac{1}{T} \sum_{t=1}^{T} 
    \frac{\pi_{t}^{\mathrm{BESS}} - \pi_{t}^{\mathrm{no\,BESS}}}{S^{\max}}.
\end{equation}
Standardizing the profit by the BESS capacity enables a meaningful comparison across systems of different scales. Since $V$ is computed as the difference between the income of the system with and without storage, it directly reflects the economic gains attributable to the presence of the BESS. Although the resulting profits may vary with the underlying generation and consumption structure, they are not directly tied to the net generation level. Instead, the benefits arise from three primary mechanisms: price arbitrage, load shifting and the avoidance of the grid fee due to self-consumption.

Next, the performance of the BESS can also be assessed from the perspective of battery utilization. A convenient metric is the fraction of hours during which the battery is actively operated, that is, either charged or discharged. Let
\[
    U_{t,h} = 
    \begin{cases}
        1, & \text{if } E_{t,h}^{\mathrm{ch}} \neq 0 \text{ or } E_{t,h}^{\mathrm{dis}} \neq 0,\\[4pt]
        0, & \text{otherwise},
    \end{cases}
\]
denote an indicator of battery activity at hour $h$ on day $t$. The utilization rate over the evaluation period is then defined as
\begin{equation}\label{eq:utilization}
    U = \frac{1}{24T} \sum_{t=1}^{T} \sum_{h=1}^{24} U_{t,h}.
\end{equation}

This measure captures the proportion of hours in which the battery performs active energy exchanges. Systems with longer charging times (larger $H$) generally exhibit higher utilization, as more hours are required to complete charging or discharging cycles. Conversely, a BESS with $H = 1\,\mathrm{h}$ tends to operate less frequently, since it can complete full cycles in shorter time intervals.

Finally, when the system includes consumption, it is useful to evaluate whether BESS reduces the amount of electricity purchased from the grid. 
The part of consumption covered by purchases from the grid in the system with BESS is given by
\begin{equation}\label{eq:PC_BESS}
    PC^{\mathrm{BESS}} = \frac{1}{T}\sum_{t=1}^{T}\sum_{h=1}^{24} 
    |X_{t,h}^{(-)}|,
\end{equation}
whereas for the system without BESS it becomes
\begin{equation}\label{eq:PC_noBESS}
    PC^{\mathrm{no\,BESS}} 
    = \frac{1}{T}\sum_{t=1}^{T}\sum_{h=1}^{24} 
      |NG_{t,h}|\mathbf{1}_{\{NG_{t,h}<0\}}.
\end{equation}
The additional purchases from the grid resulting from the integration of the BESS are measured as
\begin{equation}\label{eq:APC}
    APC = PC^{\mathrm{BESS}} - PC^{\mathrm{no\,BESS}}.
\end{equation}

A positive value ($APC > 0$) indicates that the system purchases more electricity when BESS is used. This increase reflects energy losses due to reduced charging and discharging efficiencies as well as additional purchases required for price arbitrage strategies. Conversely, a negative value ($APC < 0$) implies that the BESS reduces grid purchases without altering the underlying consumption profile. This effect can be particularly beneficial in systems with high grid fees, where reducing purchased energy significantly lowers operating costs.

\section{Results}\label{sec:results}

\subsection{System specification}\label{sub:sec:system:spec}

In this research, the following BESS specifications are considered. First, the net energy capacity is set to $S^{\max} = 1\,\mathrm{MWh}$. This assumption is not restrictive, as the profit measure $\pi$ is expressed per 1~MWh of energy capacity, thereby removing dependence on the absolute BESS size. When systems with generation and/or consumption are evaluated, the battery capacity is kept fixed while the levels of supplied or demanded energy vary to reflect different system configurations.

Next, four different charging times are examined: $H = 1\,\mathrm{h}, 2\,\mathrm{h}, 3\,\mathrm{h}, 4\,\mathrm{h}$, representing typical values available on the commercial market. Finally, the charging and discharging efficiency is set to $90\%$, which corresponds to 80\% round-trip efficiency often assumed in the literature \citep{bass:etal:18,ade:etal:2018,mer:etal:23}.


When the system includes consumption units, they are assumed to represent a company operating a single shift with a flat demand profile. Hourly electricity consumption during working days is defined as
\[
    C_{t,h} =
    \begin{cases}
        c, & \text{if } h \in \{7,8,\ldots,18\},\\[4pt]
        0, & \text{otherwise}.
    \end{cases}
\]
with daily consumption of $C=12c$. The unit is additionally assumed not to consume electricity during weekends. Various levels of $c$ are examined in this study, ranging from $0.1\,\mathrm{MWh}$ to $1\,\mathrm{MWh}$, representing different relationships between BESS net energy capacity and local consumption.

Similarly, the generation is assumed to be constant over the entire day, with $G_{t,h} = g$, yielding a total daily generation of $G = 24g$. Hence, the ratio of total generation to total consumption (when both are present) equals $G/C = 24g / 12c = 2g/c$. The factor 2 reflects the fact that electricity is generated continuously throughout the day, whereas demand is positive only during the 12 working hours.

These stylized consumption and generation schedules are deterministic and have a simplified structure, as this study aims to describe general relationships between self-consumption, energy purchases, and energy sales. This simplifying assumption allows us to isolate the effect of grid fees and market prices on BESS operation without introducing additional uncertainty related to demand and generation variability. In real-life applications, both generation and consumption are more divers \citep{tiem:etal:20} and should be modeled as stochastic processes \citep{sel:etal:26}.

It is assumed that BESS participates in the DA market; therefore, scheduling decisions rely on forecasted prices. At the same time, the final performance evaluation is based on actual prices. In this article, BESS is assumed to know electricity prices in advance and to use this information when making operational decisions. This assumption is common in the literature (see \cite{staf:rus:16} and \cite{mer:etal:23} for a discussion), as it allows researchers to assess ex post the impact of market price dynamics on BESS profitability without exposing to the risk of forecast selection. Moreover, it provides an estimate of the upper-bound of the value of BESS, which can be used as the benchmark for real-life applications.

The electricity prices used in the analysis are hourly day-ahead  prices for the  Germany–Luxembourg bidding zone, sourced from the ENTSO-E transparency platform. They span three years of data between 01.01.2022 and 31.12.2024. The data is adjusted for daylight saving time transitions: missing hours during the spring shift to CEST are imputed using the average of adjacent hours, while duplicated hours during the autumn return to CET are replaced with their mean.

Finally, to reflect the changing regulatory framework, different levels of distribution fees are considered. A grid fee of $0\,\mathrm{ct/kWh}$ represents a hypothetical situation in which network charges do not depend on the quantity of electricity purchased. As in real life applications, the part of the payments for usage of grid is variable, positive values of $F$ are more plausible.  In 2024, grid fees in Germany averaged approximately $4\,\mathrm{ct/kWh}$ for energy-intensive industries, $9\,\mathrm{ct/kWh}$ for other commercial consumers, and around $11\,\mathrm{ct/kWh}$ for households. 
In 2026, further reductions in grid fees are expected, with the aim of lowering electricity purchase costs and supporting the competitiveness of consumers and distributed energy resources. To capture both current and anticipated market conditions, this study analyzes results for grid fees ranging from 
\[
    F = 0, 1, \ldots, 12 \ \mathrm{ct/kWh}.
\]
These values span the relevant regulatory range and allow the assessment of BESS performance under various policy scenarios.

\subsection{BESS without generation or consumption}

Let us first consider the system that includes only the BESS. In this case, it is assumed that $g = c = 0$. The system therefore generates additional income solely through price arbitrage: purchasing electricity (charging the battery) when prices are low and selling electricity (discharging the battery) when prices are high. The performance of the BESS is summarized in Table~\ref{tab:only BESS}, which reports the added value and utilization rate of the battery for different charging times and levels of the grid fee. 

The results show that the added value, $V$, decreases as the charging time $H$ increases. When the grid fee is zero, extending the charging time from $H=1\,\mathrm{h}$ to $H=4\,\mathrm{h}$ reduces the addiction income by 28.2\%. At the same time, utilization triples, rising from 14.98\% to 46.88\%. The relative decline in financial gains from installing BESS becomes even more pronounced at higher fee levels.

The grid fee itself has a substantial impact on both profitability and utilization. For a BESS with $H = 1\,\mathrm{h}$, increasing the fee from 0 to 12~ct/kWh reduces added value by 80.1\%, from 130.07~EUR to 25.94~EUR, and lowers utilization from 14.98\% to 2.74\%, leaving the battery almost unused. Under the current fee level for regular entreprenEUR ($F = 9$~ct/kWh), the profitability of investing in a BESS is questionable: the annual profit of 1,401.4~EUR (corresponding to approximately 142,014~EUR over ten years) is insufficient to cover typical investment costs. If the fee is reduced, profitability improves substantially, rising to 20{,}684.6~EUR and 26,732.6~EUR per year for fee levels of 6 and 4~ct/kWh, respectively.

\begin{table}
\centering
\caption{Performance of a stand-alone BESS in the absence of generation and consumption units.}
\label{tab:only BESS}
 \begin{tblr}{
  colspec={c|cccc},
  cell{1}{2} = {c=4}{c},
  cell{3,10}{2} = {c=4}{c}
}
\hline
Grid fee & Time to charge (H) & & & \\
\cline{2-5}
(ct/kWh) &  1h & 2h& 3h& 4h \\
\hline
 &Added value, $V$ (EUR) & & &\\
\hline
0 & 130.07 & 117.10 &  104.28 &  92.56\\
2 &  96.29 &  85.60 &  75.68  &  66.96\\
4 &  73.24 &  64.63 &  56.88  &  50.15\\
6 &  56.67 &  49.37 &  43.05  &  37.76\\
9 &  38.36 &  33.00 &  28.48  &  24.83\\
12&  25.94 &  22.18 &  18.99  &   16.50\\
\hline
 &BESS utilization, $U$ (\%) & & &\\
\hline
0  &  14.98 &  27.83 &   38.80 &   46.88\\
2  &  10.56 &  19.15 &   25.77 &   30.37\\
4  &  7.52  &  12.97 &   17.91 &   20.95\\
6  &  5.64  &   9.73 &   13.41 &   15.48\\
9  &  4.04  &   6.52 &    8.87 &    9.84\\
12 &  2.74  &   4.11 &    5.73 &    6.15\\
\hline
\end{tblr}

\vspace{0.3cm}
\small{Notice: System with no consumption or generation ($C_{t,h} = G_{t,h}=0$).  BESS specification: $e=90\%$, $S^{max} = 1MWh$, $D=2$. }
\end{table}

\subsection{BESS with a consumption unit}

Accompanying  BESS with a consumption unit provides additional sources of income. Let us first consider BESS located next to a company with the demand profile described in Section~\ref{sub:sec:system:spec}. In this analysis, four potential hourly consumption levels are considered: $c = 0.1, 0.2, 0.5,$ and $1\,\mathrm{MWh}$, which correspond to daily demand levels of $C = 1.2, 2.4, 6,$ and $12\,\mathrm{MWh}$, respectively. The resulting added values for different grid fees and for two selected charging times, $H = 1\,\mathrm{h}$ and $H = 4\,\mathrm{h}$, are presented in Table~\ref{tab:BESS:cons}. Let us recall here, that for all considered levels of consumption, the battery net energy capacity is $1\,\mathrm{MWh}$. Hence, the larger the value of $c$, the smaller the battery becomes relative to the total consumption level.

First, it can be observed that when the grid fee is zero, the added value from installing a BESS is identical to the profit obtained by a stand-alone battery without any accompanying units (see Table~\ref{tab:only BESS}). In the absence of any cost associated with purchasing electricity from the grid, the two operational strategies become equivalent:  
(i) discharging the battery to directly cover the local demand, or  
(ii) discharging the battery to sell electricity to the grid and subsequently purchasing the required energy from the grid to meet the demand.  
Since both approaches yield the same financial outcome when $F = 0$, the presence of a consumption unit does not generate additional gains beyond those obtained from pure price arbitrage.

When the added values of integrating BESS into the system are examined for non-zero grid fees, it becomes evident that BESS generates additional gains beyond those obtained from pure price arbitrage. For instance, when $F = 9\,\mathrm{ct/kWh}$, a BESS with $H = 1\,\mathrm{h}$ yields an income of 48.29--62.69~EUR per day, depending on the total level of consumption, whereas a stand-alone battery earns only 38.36~EUR. These additional gains arise primarily from consumption shifting, which enables the system to purchase electricity at lower prices while incurring only a minor increase in grid usage costs.

Finally, the results indicate that the profitability of the BESS increases with the relative size of consumption. For example, for $F=9\,\mathrm{ct/kWh}$ the difference in daily added values between systems with $c = 0.1\,\mathrm{MWh}$ and $c = 1\,\mathrm{MWh}$ reaches 14.40~EUR for $H = 1\,\mathrm{h}$ and 5.86~EUR for $H = 4\,\mathrm{h}$. This increase reflects the more efficient utilization of the battery, particularly due to the greater potential for demand shifting in systems with higher consumption levels.

\begin{table}
\centering
\caption{Average daily added value, $V$ (EUR), of a BESS accompanying a consumption unit. }
\label{tab:BESS:cons}
 \begin{tblr}{
  colspec={c|cccc|ccc},
  cell{3}{2,6} = {c=4}{c},
  cell{1}{2} = {c=8}{c},
}
\hline
 & Hourly consumption ($c$, MWh) & & & &  & & & \\
 \cline{2-9}
(ct/kWh) &  0.1 & 0.2 & 0.5 & 1.0 &  0.1 & 0.2 & 0.5 & 1.0  \\
\cline{2-9}
Grid fee &  $H=1h$ & & & &  $H=4h$ & & & \\
\hline
 0 & 130.07 & 130.07 & 130.07 & 130.07 & 92.56  & 92.56  & 92.56  & 92.56  \\
 2 & 98.16  & 99.85  & 103.34 & 105.51 & 70.16  & 72.63  & 73.19  & 73.19  \\
 4 & 77.27  & 80.26  & 85.34  & 88.09  & 56.01  & 59.62  & 60.40  & 60.40  \\
 6 & 63.04  & 66.97  & 72.76  & 75.78  & 46.11  & 50.51  & 51.41  & 51.41  \\
 9 & 48.29  & 53.10  & 59.45  & 62.69  & 36.42  & 41.32  & 42.28  & 42.28  \\
12 & 38.63  & 43.78  & 50.13  & 53.46  & 30.20  & 35.09  & 36.03  & 36.03  \\
\hline

\end{tblr}

\vspace{0.3cm}
\small{Notice: System with no generation ($G_{t,h}=0$). BESS specification: $e=90\%$, $S^{max} = 1MWh$, $D=2$.}
\end{table}

\subsection{BESS with a generation unit}

Next, we consider BESS accompanying a generation unit. In this configuration, the battery can be used by the producer either to perform price arbitrage or to temporally shift trade. In the latter case, the utility is able to sell electricity under more favorable market conditions and thus reduce exposure to price risk. 
The gains from installing BESS are reported in Table~\ref{tab:BESS:gen}. It presents the added value, $V$, for different levels of grid fees and hourly generation $g = 0.05, 0.1, 0.2,$ and $0.5\,\mathrm{MWh}$, which correspond to a daily generation of $G = 1.2, 2.4, 4.8,$ and $12\,\mathrm{MWh}$, respectively. 
As before, the net energy capacity of the battery is fixed at 1\,MWh. Thus, higher values of $G$ imply that the relative capacity of the BESS becomes smaller compared to the level of the load.

Similarly to the case of BESS coupled with a consumption unit, when $F = 0$ the added value of the BESS is equal to that of a stand-alone battery. However, as the grid fee increases, the differences between these two configurations become increasingly pronounced. For a low level of generation, $g = 0.05\,\mathrm{MWh}$, and a short charging time, $H = 1\,\mathrm{h}$, the profits from installing  BESS remain above 65~EUR even for $F = 9\,\mathrm{ct/kWh}$. In this case, they exceed those of the stand-alone battery by almost 30~EUR per day.

The gains from using the battery further increase as total generation rises. For example, for $F = 9\,\mathrm{ct/kWh}$ and $g = 0.1\,\mathrm{MWh}$, the daily profit reaches 88.38~EUR. In an extreme case, when $g = 0.5\,\mathrm{MWh}$, the additional revenue from using the BESS approaches 125.17~EUR per day, corresponding to 45,687.1~EUR per year. Moreover, for high generation level, as compared to the energy capacity of the battery,  the profit is only weakly dependent on the level of the grid fee. These results indicate that, as the grid fee increases, the income generated from temporal load shifting (which does not incur grid usage costs) may outweigh the losses associated with reduced price-arbitrage opportunities.

Finally, when comparing the added value of a BESS located close to a consumption unit with that of a BESS co-located with a generation facility, it becomes evident that combining storage with generation is the more profitable strategy. Even when the level of electricity production is relatively low -- for example, $g = 0.1\,\mathrm{MWh}$, corresponding to a total daily generation of 1.2~MWh -- the resulting gains exceed those of systems with substantially higher daily consumption such as $C = 2.4$ or 6~MWh, especially for high levels of $F$. The higher additional income from a generation-coupled BESS stems from its ability to shift load without incurring grid fees. In contrast, consumption shifting does not reduce grid usage costs, as the amount of energy purchased from the market is fixed by the demand level and cannot be lowered through temporal reallocation.

\begin{table}
\centering
\caption{Average daily added value, $V$ (EUR), of a BESS accompanying a generation unit. }
\label{tab:BESS:gen}
 \begin{tblr}{
  colspec={c|cccc|cccc},
   cell{3}{2,6} = {c=4}{c},
  cell{1}{2} = {c=8}{c},
}
\hline
Grid fee & Hourly generation ($g$, MWh) & & & &  & & & \\
 \cline{2-9}
 (ct/kWh) &  0.05 & 0.1 & 0.2 & 0.5 &  0.05 & 0.1 & 0.2 & 0.5  \\
\cline{2-9}
 &  $H=1h$ & & & &  $H=4h$ & & & \\
\hline
 0 & 130.07 & 130.07 & 130.07 & 130.07 & 92.56 & 92.56 & 92.56 & 92.56 \\
 2 & 101.92 & 107.31 & 116.07 & 125.68 & 74.10 & 80.16 & 88.66 & 92.56 \\
 4 & 86.08  & 97.21  & 113.10 & 125.29 & 64.20 & 74.67 & 87.64 & 92.56 \\
 6 & 76.57  & 92.07  & 112.10 & 125.21 & 57.93 & 71.67 & 87.27 & 92.56 \\
 9 & 68.25  & 88.38  & 111.64 & 125.17 & 52.82 & 69.61 & 87.09 & 92.56 \\
12 & 64.37  & 86.96  & 111.47 & 125.14 & 50.79 & 68.92 & 87.03 & 92.56 \\
\hline
\end{tblr}

\vspace{0.3cm}
\small{Notice: System with no consumption ($C_{t,h}=0$). BESS specification: $e=90\%$, $S^{max} = 1MWh$; decision horizon $D=2$. }
\end{table}

\subsection{BESS with both a consumption and a generation unit}

Finally, a BESS accompanying both a generation and a consumption unit is analyzed. As before, the hourly consumption level is assumed to range between 0.1 and 1~MWh. The level of generation is expressed relative to consumption and described by the ratio $g/c$. When the ratio equals 0\%, no generation is present and the results correspond to the configuration with a BESS and a consumption unit only. Larger values of the ratio $g/c=10\%, 20\%, 50\%$ or 100\% represent different system scenarios. Since the net capacity of the battery is fixed at 1~MWh, an increase in consumption and/or generation implies that the relative capacity of the BESS becomes smaller compared to the size of the accompanying units.

\begin{figure}
\centering
\begin{subfigure}{0.49\textwidth}
    \includegraphics[width=\textwidth]{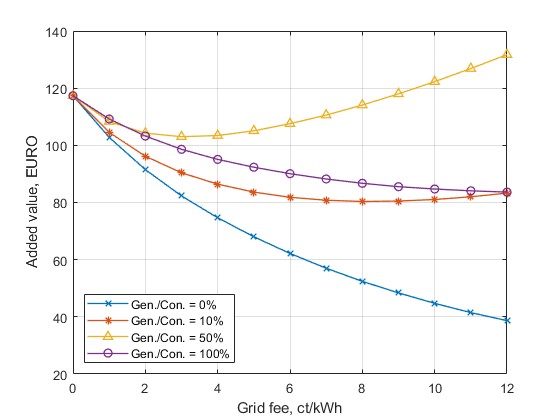}
    \caption{Added value, $V$}
    \label{fig:profit:a}
\end{subfigure}
\hfill
\begin{subfigure}{0.49\textwidth}
    \includegraphics[width=\textwidth]{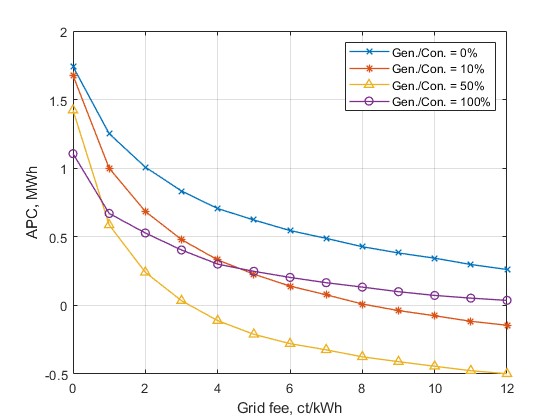}
    \caption{Additional purchase from the grid, $APC$}
    \label{fig:profit:b}
\end{subfigure}   
\caption{Value added per 1MWh of energy capacity ($V$) and dditional purchase from the grid ($APC$) for different hourly generation levels (expressed as a percentage of the hourly consumption); BESS specification: $e=90\%$, $S^{max} = 1\,\mathrm{MWh}$, $H=1\,\mathrm{h}$, $D=2$; hourly consumption $c = 0.5\,\mathrm{MWh}$.}
\label{fig:profit}
\end{figure}

{We first consider a system with $c = 0.1\,\mathrm{MWh}$ and the charging time $H = 1\,\mathrm{h}$. The added value across different grid fee levels and four considered values of $g/c$ is presented in the left panel of Fig.~\ref{fig:profit}. In the absence of generation, $g/c=0\%$,  the gains from installing the BESS decrease as the grid fee increases. However, as the ratio of generation to consumption rises, the added value no longer follows a monotonic pattern. Initially, it declines because higher grid fees reduce price-arbitrage opportunities. Beyond a certain threshold, however, the gains begin to increase, as the BESS enables a reduction in distribution costs by shifting both generation and consumption in time. }

{This behavior is further confirmed by the level of additional purchases, depicted in Fig.~\ref{fig:profit:b}, which measures the difference in electricity purchased from the grid between systems with and without the BESS, as defined in~\eqref{eq:APC}. For small values of $F$, the measure $APC$ is positive and exceeds 1~MWh, indicating a significant role of price arbitrage in generating profits. As the distribution fee becomes more pronounced, load shifting and self-consumption gain importance, resulting in a reduction of purchases and a decline of $APC$ below zero. When $F = 12\,\mathrm{ct/kWh}$, $APC$ falls to approximately -0.5~MWh for systems with large relative generation. Even for moderate generation levels with $g/c = 10\%$, $APC$ remains negative for grid fees above 8~ct/kWh.}

The performance of the BESS is finally compared across different levels of consumption, generation-to-consumption ratios, and two selected grid fee levels: $F = 4\,\mathrm{ct/kWh}$ and $F = 9\,\mathrm{ct/kWh}$, which correspond to current distribution fee conditions. The added value generated by the BESS, along with the additional purchases, is presented in Table~\ref{tab:BESS:con:gen}. These results provide a broader perspective on the profitability of battery integration across various system configurations.

The results reported in the upper block of Table~\ref{tab:BESS:con:gen} confirm a non-linear relationship between BESS value and the level of grid fees. For low levels of consumption ($c = 0.1\,\mathrm{MWh}$), moderate generation ($g/c=50\%$), higher distribution fees reduce the financial gains from installing energy storage from $84.78\,\mathrm{EUR}$ to $69.14\,\mathrm{EUR}$ when $F = 4\,\mathrm{ct/kWh}$ and $F = 9\,\mathrm{ct/kWh}$, respectively. This effect is mainly driven by limited arbitrage opportunities and a small potential for reducing transmission costs through load shifting.
By contrast, when BESS operates alongside higher levels of consumption ($c = 0.5$ or $1\,\mathrm{MWh}$), load shifting becomes the dominant source of value. In this case, higher distribution fees increase the economic benefit of BESS by 13.4 EUR and 17.49 EUR for $c=0.5$ and $c=1$, respectively. Larger generation enables the storage to increase significantly self-consumption and reduces transmission charges. When generation is sufficient to fully cover consumption ($g/c = 100\%$), the role of BESS in increasing self-consumption disappears. Consequently, the battery no longer contributes to lowering transmission costs, and its added value declines.

The accessibility and profitability of load shifting are illustrated in the second block of Table~\ref{tab:BESS:con:gen}, which reports the values of $APC$ across different system configurations. When load shifting reduces electricity purchases from the grid, the value of $APC$ becomes negative. The results show that the lowest values of $APC$ correspond to the highest added values of BESS, highlighting the important role of self-consumption in generating profits. 

Finally, the added value of BESS indicates that the profitability of storage depends strongly on the relative size of the battery compared with the levels of consumption and generation. For $F = 4\,\mathrm{ct/kWh}$, the added value ranges from 77.27 to 126.67~EUR, while for $F = 9\,\mathrm{ct/kWh}$ it ranges from 48.29 to 144.16~EUR. The sensitivity to system configuration becomes more pronounced as the distribution fee increases. 
When $F = 9\,\mathrm{ct/kWh}$, the largest financial gains are observed when battery capacity is relatively small compared with the consumption level ($c = 1\,\mathrm{MWh}$) and when the generation-to-consumption ratio equals 50\%. Conversely, when consumption is small relative to BESS capacity ($c = 0.1\,\mathrm{MWh}$) and no generation is available, the storage system provides only limited profit opportunities. A similar pattern is observed for the lower grid-fee level, $F = 4\,\mathrm{ct/kWh}$.

\begin{table}
\centering
\caption{Performance of a BESS accompanying a generation and a consumption unit}
\label{tab:BESS:con:gen}
 \begin{tblr}{
  colspec={c|ccc|ccc},
  cell{1}{2,5} = {c=3}{c},
  cell{3,10}{2} = {c=6}{c},
}
\hline
            &Grid fee = 4ct/kWh &      & &Grid fee = 9ct/kWh &      &      \\
\hline
Consumption (MWh) &    0.1   & 0.5   & 1.0 &   0.1   & 0.5   & 1.0  \\
           \hline
 Gen./Con. &Added value, $V$ (EUR) &      &     & &      &   \\          
\hline
  0\%  & 77.27  & 85.34   & 88.09   & 48.29 & 59.45  & 62.69\\
 10\%  & 78.99  & 95.37   & 108.40  & 53.78 & 88.73  & 117.22\\
 30\%  & 82.16  & 111.39  & 124.94  & 63.01 & 123.92 & 142.53\\
 50\%  & 84.78  & 116.56  & 126.67  & 69.14 & 130.44 & 144.16\\
 80\%  & 87.65  & 113.61  & 121.37  & 71.87 & 121.80 & 135.60\\
100\%  & 89.14  & 109.87  & 97.53   & 71.41 & 96.53  & 99.24\\
\hline
   Gen./Con.  & $APC$ (MWh) &      &     & &      &       \\
\hline
 0\%    & 0.843  & 0.737   & 0.718   & 0.389 & 0.364  & 0.360\\
 10\%   & 0.783  & 0.400   & 0.078   & 0.301 & -0.051 & -0.350\\
 30\%   & 0.677  & -0.048  & -0.212  & 0.166 & -0.385 & -0.444\\
 50\%   & 0.604  & -0.117  & -0.212  & 0.089 & -0.394 & -0.440\\
 80\%   & 0.560  & 0.045   & -0.105  & 0.134 & -0.332 & -0.411\\
 100\%  & 0.556  & 0.330   & 0.321   & 0.204 &  0.118 & 0.112\\
 \hline
\end{tblr}

\vspace{0.3cm}
\small{Notice:  BESS specification: $e=90\%$, $S^{max} = 1MWh$, time to charge $H=1h$. }
\end{table}

\subsection{Decision horizon}

The number of days, $D$, included in the decision process, may impact the scheduling and profitability of the battery. Most of the energy price forecasting literature focuses on day-ahead predictions \citep[see][for a comprehensive review]{lag:mar:des:wer:21}, which corresponds to selection of $D=1$. On the contrary, one could consider a longer forecasting horizon that allows for using information on two or three following days in the optimization algorithm. 

To assess the impact of the choice of $D$ on BESS performance, we compare the added value of a stand-alone BESS with charging durations $H = 1\,\mathrm{h}$ and $4\,\mathrm{h}$ for $D = 1, 2,$ and $3$. The results are summarized in Table~\ref{tab:only BESS:D}. They indicate that extending the scheduling horizon from one day to two days yields substantial gains, while further increasing $D$ to three days results in only minor additional improvements. For example, in a system with no grid fees and a BESS with the charging time of 1h, added value rises by 10.80\% as $D$ increases from 1 to 2. However, the change of $D$ from 2 to 3 leads to a negligible improvement of profits of only 0.03\%. The corresponding changes in added value for $F = 9\,\mathrm{ct/kWh}$ are 51.74\% and 2.01\%, confirming the the phenomena is valid across different values of transmission fees.

This pattern suggests that most intertemporal arbitrage opportunities can already be captured within a two-day optimization horizon, while additional foresight beyond this window provides only limited incremental value.
This justifies using $D = 2$ as the baseline decision horizon, as it captures most of the economic benefits of multi-day optimization while avoiding unnecessary computational complexity and reliance on longer-term price forecasts.

\begin{table}
\centering
\caption{Performance of a stand-alone BESS in the absence of generation and consumption units.}
\label{tab:only BESS:D}
 \begin{tblr}{
  colspec={c|ccc|ccc},
  cell{1}{2} = {c=6}{c},
  cell{3}{2,5} = {c=3}{c}
}
\hline
Grid fee & Decision horizon ($D$) & &  \\
\cline{2-74}
(ct/kWh) &  1 & 2& 3 &  1 & 2& 3 \\
\cline{2-74}
&   $H=1h$ & & & $H=4h$ & & \\
\hline
0 & 117.39 & 130.07 &  130.12 & 80.64   &  92.56  & 92.59\\
2 &  82.18 &  96.29 &  96.33  & 56.06   &  66.96  & 66.96\\
4 &  58.31 &  73.24 &  73.27  & 39.58   &  50.15  & 50.20\\
6 &  41.88 &  56.67 &  57.09  & 27.99   &  37.76  & 38.21\\
9 &  25.28 &  38.36 &  39.13  & 16.49   &  24.83  & 25.64\\
12&  15.62 &  25.94 &  26.93  &  9.94   &  16.50  & 17.58\\
\hline
\end{tblr}

\vspace{0.3cm}
\small{Notice: Profit is calculated as an average daily profit from BESS per 1MWh of energy capacity.  BESS specification: $e=90\%$, $S^{max} = 1MWh$. }
\end{table}

The impact of $D$ on BESS operation may change when imperfect electricity price forecasts are used instead of perfect foresight in the optimization algorithm. Suppose that the company bases its decisions on naive predictions, in which electricity prices are assumed to remain unchanged in the following days. The forecast is therefore defined as
$$ \hat{P}_{d+D,h} = P_{d-1,h}. $$
This type of prediction does not incorporate any information about potential changes in market prices resulting from variable weather conditions or other market factors.

The resulting gains from a stand-alone BESS are presented in Table~\ref{tab:only BESS:D:naive}. A comparison with outcomes based on perfect foresight leads to several conclusions. First, the use of imperfect forecasts substantially reduces profits. For a system with zero grid fees, profits decrease by 36.50~EUR and 24.77~EUR for BESS with one- and four-hour charging times, respectively. Moreover, as transmission charges increase, the gains shrink to zero and become negative for $F = 12\,\mathrm{ct/kWh}$. 

Second, consistent with previous results, only minor differences are observed between outcomes obtained for $D = 2$ and $D = 3$. At the same time, for moderate grid fee levels, extending the decision horizon from $D = 1$ to $D = 2$ leads to a substantial increase in profits.

Finally, the results for $H = 4\,\mathrm{h}$ show that, in some cases, extending the decision horizon reduces battery value. This outcome has important practical implications, indicating that imprecise longer-term forecasts may hamper the decision-making process. Hence, it supports the use of a short, one- or two-day decision horizon.

\begin{table}
\centering
\caption{Performance of a stand-alone BESS in the absence of generation and consumption units. Naive forecasts}
\label{tab:only BESS:D:naive}
 \begin{tblr}{
  colspec={c|ccc|ccc},
  cell{1}{2} = {c=6}{c},
  cell{3}{2,5} = {c=3}{c}
}
\hline
Grid fee & Decision horizon ($D$) & &  \\
\cline{2-74}
(ct/kWh) &  1 & 2& 3 &  1 & 2& 3 \\
\cline{2-74}
&   $H=1h$ & & & $H=4h$ & & \\
\hline
0 & 80.89  &  89.14 &  89.10 & 55.87   &  65.38  & 65.49\\
2 & 48.05  &  57.07 &  57.27 & 31.87   &  38.22  & 38.32\\
4 & 26.82  &  35.68 &  35.91 & 16.95   &  20.45  & 20.63\\
6 & 13.09  &  20.11 &  20.37 &  7.19   &  8.52   &  8.65\\
9 &  1.87  &   4.94 &   5.45 &  0.41   &  -0.30  &  0.07\\
12& -1.12  &  -1.13 &  -0.76 & -1.79   &  -2.79  & -2.46\\
\hline
\end{tblr}

\vspace{0.3cm}
\small{Notice: Profit is calculated as an average daily profit from BESS per 1MWh of energy capacity.  BESS specification: $e=90\%$, $S^{max} = 1MWh$. }
\end{table}

\section{Conclusions}\label{sec:conclusions}

This article examines how distribution fees affect the operation and profitability of battery energy storage systems (BESS) participating in the day-ahead electricity market. Using a mixed-integer linear programming framework and market data from Germany, we evaluate stand-alone storage as well as BESS combined with consumption and generation units under different levels of grid fees.

The results show that grid fees play a decisive role in determining BESS value. For stand-alone storage, higher transmission charges reduce arbitrage opportunities and lower profitability. When BESS is combined with consumption or generation units, additional value arises from load shifting and increased self-consumption. In such configurations, the relationship between grid fees and BESS value becomes non-monotonic: while higher fees reduce arbitrage revenues, they simultaneously increase the economic benefit of load management and transmission cost savings.

The analysis also shows that the relative size of storage capacity compared with consumption and generation levels is a key determinant of BESS profitability. From an operational perspective, extending the optimization horizon from one day to two days significantly improves scheduling outcomes, whereas further extensions provide only marginal benefits, particularly when forecasts are imperfect.

\paragraph{Policy implications}
The findings suggest that grid tariff design can substantially influence both the profitability and operational role of energy storage. Energy-based grid fees may unintentionally discourage price arbitrage while simultaneously encouraging self-consumption and load shifting. As a result, tariff structures can shape how storage contributes to system flexibility and renewable integration. Policymakers should therefore consider the interaction between grid fees, distributed generation, and storage operation when designing transmission and distribution tariffs. In particular, regulatory frameworks that ensure cost-reflective but technology-neutral grid charges may help avoid distortions in storage investment and operation decisions.

Future research could extend the analysis by incorporating stochastic demand and generation profiles, battery degradation costs, participation in ancillary service markets, and alternative grid tariff structures.

\bibliography{ref}

\end{document}